# Optical galleries of dielectric rings: antipode to whispering gallery modes

A. P. Chetverikova[1], M. S. Sidorenko[1], K. B. Samusev[1,2], M. F. Limonov[1,2], N. S. Solodovchenko[1]

[1]*School of Physics and Engineering, ITMO University, 191002, St. Petersburg, Russia*
[2]*Ioffe Institute, St. Petersburg 194021, Russia*

A general picture of photonic eigenmodes of dielectric rings with a rectangular cross section is presented, which is fundamentally different from those of whispering gallery modes of the disc. The optical spectrum of a rectangular dielectric ring consists of an infinite set of individual galleries, each starting with broad transverse radial or axial Fabry–Pérot-like resonances due to two pairs of side faces. Each gallery continues with a set of equidistant longitudinal modes with exponentially increasing quality factors. Transverse radial and axial modes exhibit strict periodicity of line contours in the Fano-Lorentz-Fano-Lorentz sequence due to the periodicity of cylindrical harmonics. Theoretical and experimental results are in perfect agreement.

**Introduction.** The optical spectra of axisymmetric dielectric resonators such as disks, rings, spheres, and toroids are the focus of researchers' attention. A surge of attention is associated with the observation of induced magnetic resonances in high-index dielectric nanoparticles [1,2]. Due to optical Mie-resonances accompanied by strong localization of both electric and magnetic fields, dielectric structures are expected to replace various plasmonic components in a number of potential applications, since dielectrics do not suffer from conduction losses [3].

A complete picture of light scattering by dielectric structures cannot be imagined without the contribution of resonances of a different nature, such as Fabry–Pérot-like resonances and whispering gallery modes (WGMs). It is Fabry–Pérot-like resonances, together with the Mie-like resonances, that lead to the formation of bound states in the continuum in dielectric cylinders [4] and rings [5]. The spectra of light scattering by a dielectric cylinder have been studied in detail [6,7], which cannot be said about the situation with dielectric ring resonators (RRs). The RR is one of the main building blocks of advanced integrated optical circuits [8,9]. In addition to many applications, dielectric RRs are actively used for basic research, including the concept of an optical isolator based on resonance splitting in a silicon RR [10], an observation of the multidimensional Purcell effect in an ytterbium-doped RR [11], and a generation of quantum-correlated photon pairs using electrically tunable RRs [12].

Oddly enough, the extensive literature on RRs indicates that the efforts of the developers were aimed at creating RRs operating similar to disk resonators, that is, in a WGM regime [9,13,14]. WGMs are confined within the structure by total reflections from the rim with the proper phase condition after circling along the whole resonator [15]. However, the presence of two side walls in RRs leads to a strong light confinement and the difference from a disk resonator becomes fundamental. We have just demonstrated that the low-frequency scattering spectrum of a thin-height RR splits into separate photonic mode galleries, each of which starts with a broad radial Fabry–Pérot-like resonance defined by the two sidewalls and continues with a limited set of equidistant longitudinal modes with an exponentially increasing *Q* factors [16]. Moreover, Fano resonances [17,18] were discovered in the scattering spectra of dielectric RRs, and a strict periodicity of the contours of broad Fabry–Pérot-like resonances according to the Lorentz-Fano-Lorentz-Fano sequence was established, which has not yet been observed in any other structures. It should be noted that in Ref. [16] only thin RRs were experimentally studied, their height did not change; therefore, Fabry–Pérot-like resonances, which should arise due to the parallel upper and lower walls of the ring, were not observed and the complete picture of photonic resonances was not presented.

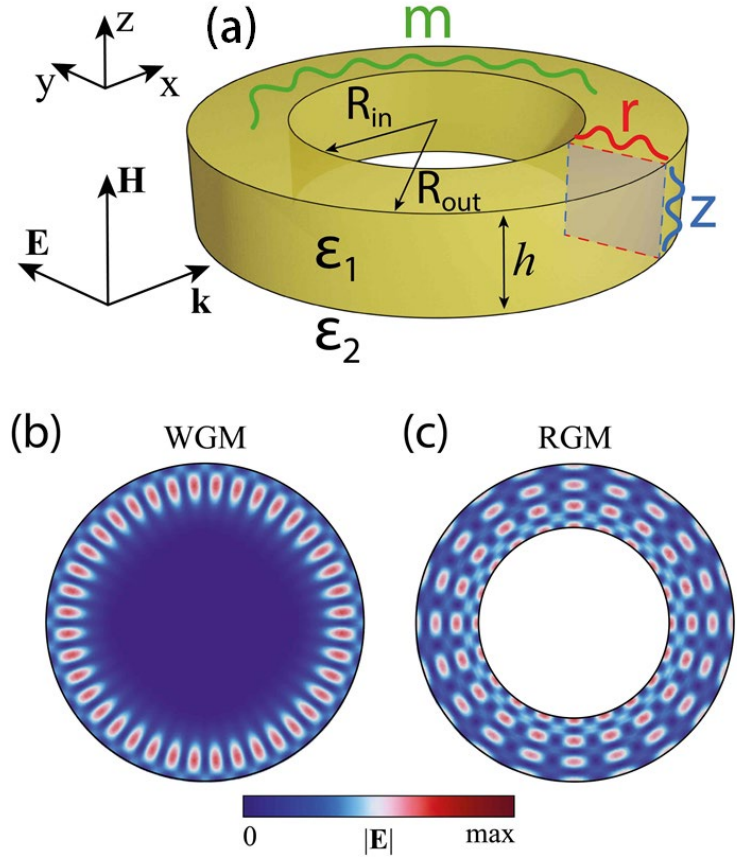

**Fig. 1.** (a) TE-polarized waves incident on a dielectric RR with permittivity $\varepsilon_1$, inner radius $R_{in}$, outer radius $R_{out}$, and height $h$ placed in vacuum ($\varepsilon_2 = 1$). The azimuthal ($m$), radial ($r$) and axial ($z$) mode indices are indicated. Schematically shown: longitudinal azimuthal and two transverse Fabry-Perot-like resonances. Examples of calculated patterns of the electric field amplitude $|\mathbf{E}|$ for the WGM (20,1,0) of a dielectric disk (b) and for the RGM (10,3,0) of a dielectric ring (c). The color scale corresponds to the field amplitude $|\mathbf{E}|$.

Here, the general picture of light scattering by dielectric RRs of variable sizes is established experimentally and theoretically, when the electromagnetic properties are determined by all four side walls of the ring. It is shown that as the height of the ring increases, in addition to the radial photonic galleries, axial galleries are added, each of which begins with a broad Fabry-Perot-like resonance between the upper and lower walls of the ring and continues with a set of equidistant longitudinal modes. By analogy with the term WGM, we will further refer to these series as "ring gallery modes" (RGM). The fundamental difference between the WGM and the RGM is determined by the mechanisms of the eigenmodes appearance in cylindrical and ring resonators. WGMs are confined within the structure by total reflections from the rim with the proper phase condition after circling along the whole resonator [19]. WGMs are formed by equidistant set of resonances with similar amplitudes and Q-factors for adjacent lines, which occupy a small part of the resonator volume along the side boundary, Fig. 1(b). On the contrary, RGMs have a different nature: each gallery is formed by two types of modes (one transverse and a set of longitudinal modes with an exponentially increasing $Q$-factor) and fills the entire volume of the RR, Fig. 1(c).

With this article and its title, we want to point out that there is a stereotype that any ring can be called a WGM resonator. We emphasize the fact that the ring resonator and the WGM disk resonator differ not only in their geometric shape, but also in their optical properties and practical applications.

Radial and axial galleries of dielectric rings. In this work, calculations of scattering spectra, resonant frequencies of eigenmodes (eigenvalues), and field distributions (corresponding eigenfunctions) are carried out (COMSOL software) for RRs with dielectric parameters corresponding to experimental samples, the average value of the dielectric constant $\varepsilon=43$ for a frequency range of 1-10 GHz [16]). To generalize the results to different frequency regions, the relative geometrical parameters of the RR were used. The aspect ratio $R_{in}/R_{out}$ was fixed at 0.81 with $R_{out}/h$ varying from 3.5 to 15. The key result was the interpretation of the lines in scattering spectra, that is, the determination of ordered triple (*m, r, z*) indices characterizing their eigenfunctions, with azimuthal (*m*), radial (*r*) and axial (*z*) mode indices. According to the standard nomenclature, the RR modes are denoted as $TE_{mrz}$ and $TM_{mrz}$, however, for most resonances, the polarization is hybrid [20], therefore we omit the notations TE and TM, leaving only the indices.

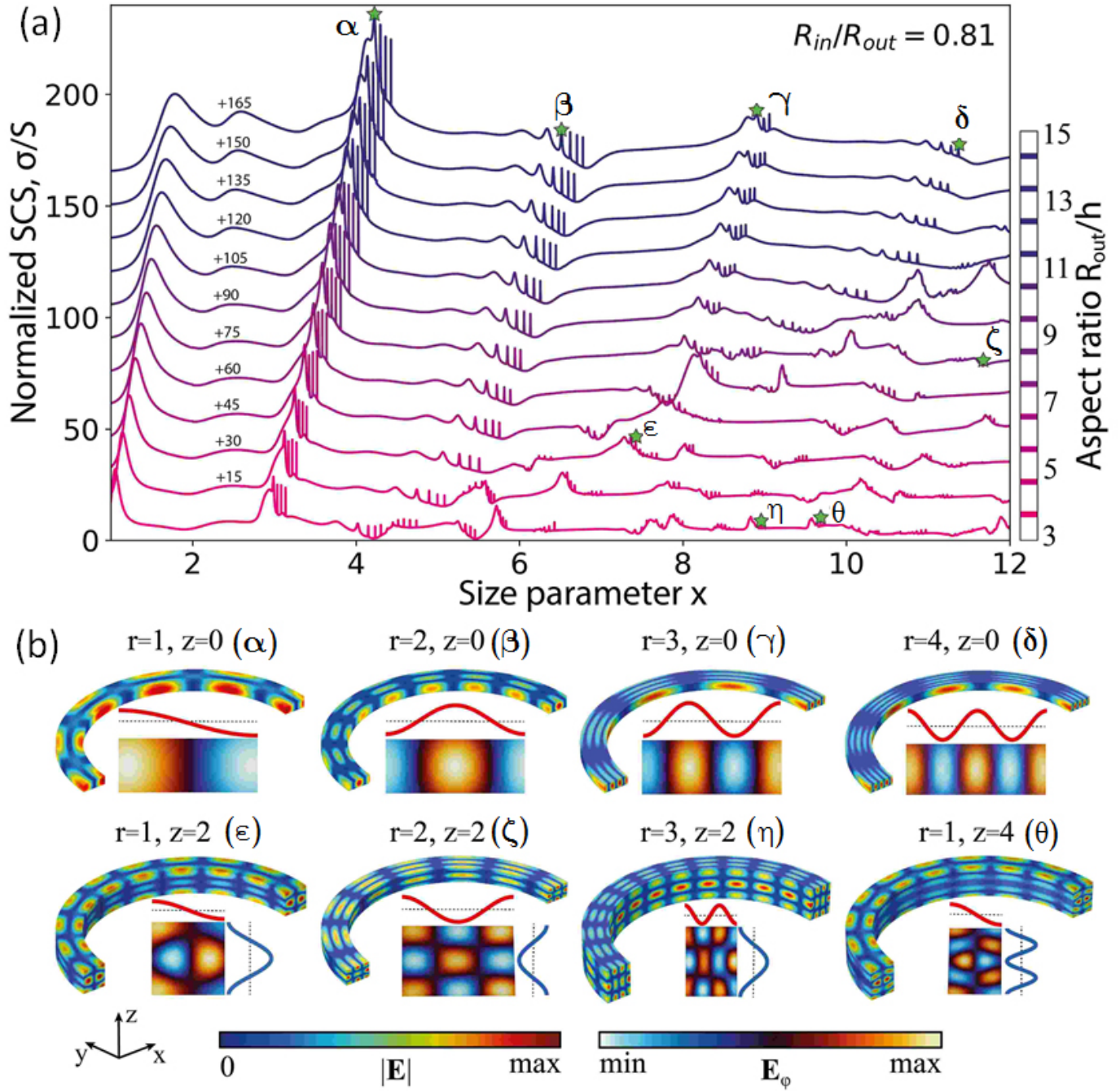

**Fig. 2.** (a) Total normalized SCS spectra of dielectric RR for various aspect ratio $R_{out}/h$ indicated to the right of the figure. The spectra are shifted vertically by the values indicated in the figure on the left. The size parameter $x = R_{out}\omega/c$. (b) Calculated field distributions in the dielectric ring for resonances marked in (a). The amplitude distributions of electric field $|E|$ are shown on half of the ring and azimuth component of electric field $E_\varphi$ on a larger scale in the cross section. The red and blue lines represent a simplified visual distribution of the field. Dashed lines correspond to zero $E_\varphi$. The radial and axial mode indices $(r, z)$ are indicated for each resonance. The two color scales correspond to the field amplitude $|E|$ and $E_\varphi$. TE polarization, $R_{in}/R_{out}=0.81$, $\varepsilon=43$, $\varepsilon_{vac}=1$.

Figure 2, which shows the calculated total normalized SCS (Scattering Cross Section) spectra and the field distribution over the RR at eight most characteristic resonant frequencies, makes it possible to trace the genesis of the photonic modes and understand the nature of the scattering spectra. The spectrum begins with an intense low frequency dipole resonance at x~1.8. Further in the spectrum, four photonic galleries are observed, with the first and third starting from a broad transverse Fabry-Pérot-like band with a quasi-Lorentz contour (at the frequencies of x~4 and ~9), and the second and fourth galleries starting from a pronounced Fano contour of the transverse Fabry-Pérot-like resonance (at frequencies of x~6.5 and ~11). In each radial and axial gallery, an intense transverse band is followed by a sequence of equidistant narrow azimuthal resonances with increasing index $m$.

When interpreting the spectra, it is important to correctly determine the indices ($m, r, z$). The index corresponds to the total number of waves that fit one round resonant trip. For azimuthal index $m$, this path is equal to the length of the ring along the midline $2\pi(R_{in} + R_{out})/2$, for radial index $r$ it is twice the width of the ring $2(R_{out} - R_{in})$, for axial index $z$ it is twice the height of the ring $2h$. In this paper, the main attention is paid to the analysis of axial galleries. By definition, for axial galleries, the index $z$ is determined by the number of half-waves corresponding to the height of the ring. As the field distribution patterns in Fig. 2(b) demonstrate, for four radial Fabry-Perot-like resonances (α-δ), the field practically does not change along the ring height due to the fact that in this spectral range the height of the ring with aspect ratio $R_{out}/h$ = 14.4 is less than half the wavelength, taking into account the dielectric constant of the material $\varepsilon$ =43. Therefore, we assign to these resonances an axial index $z=0$, while the radial index $r$ varies from 1 to 4.

New photonic axial galleries appear in the calculated spectra as the ring height increases, when half the wavelength becomes a multiple of the increasing height. As can be seen from Figs.2(b)(ε, η), a single maximum of $|E|$ is observed in the middle of the side wall, and the field decreases to zero at the edges of the side wall, which corresponds to a whole wave on half of the round resonant trip, that is, the axial index $z$ = 2. As the height of the ring increases, axial galleries with increasing index $z$ = 2 and 4 appear in succession in the observed spectral range. The color map in Fig. S1 (See Supplementary material, SM) shows the linear dependence of the wavelength on the resonator height. Therefore these are four axial galleries, the spectral position of which is directly determined by the height of the ring. This result directly confirms the nature of such photonic modes, which are determined by the axial Fabry-Pérot-like resonance between parallel lower and upper walls of the dielectric RR. In contrast, the four radial galleries shift only slightly as the height of the ring increases, as clearly seen in the behavior of the first gallery ($r=1, z=0$), Fig. 2(a).

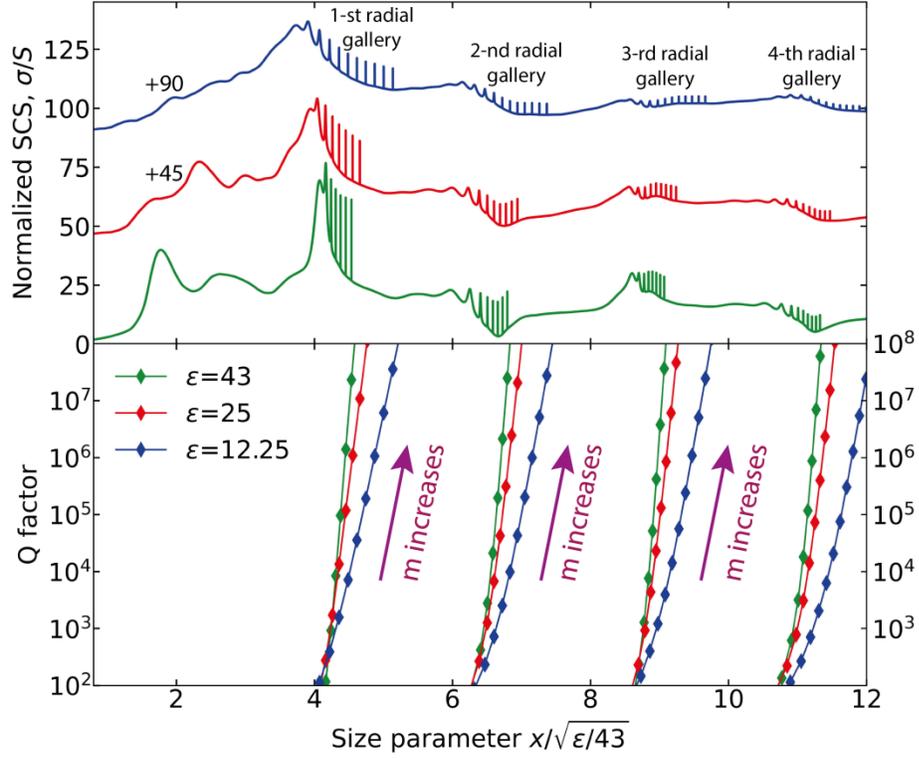

**Fig. 3.** (a) Total normalized SCS spectra of dielectric RR at three values of $\varepsilon=\mathrm{Re}(\varepsilon)+0\cdot\mathrm{Im}(\varepsilon)$ indicated in the figure. The spectra are shifted vertically by the values shown. (b) Q-factor of the longitudinal modes of the first four radial galleries depending on the rings dielectric permittivity $\varepsilon$. The colors correspond to the colors of the spectra in panel (a). TE polarization, $R_{in}/R_{out}=0.81$, $R_{out}/h = 15$.

One of the fundamental differences between RGM and WGM is the behavior of the *Q*-factor of azimuthal modes. In contrast to WGM with a practically constant *Q*-factor of resonances with a close index *m*, for RGM there is an exponential increase in the *Q*-factor with increasing *m* and this is observed in each individual gallery, as illustrated in Fig.3.

**Scattering spectrum decomposition in terms of azimuthal harmonics.** Here we demonstrate a symmetry description of the SCS spectrum of a RR using an azimuthal harmonic decomposition. The azimuthal harmonics are analogous to the Mie coefficients, which can be analyzed using the Fano formula [17, 18]. Determination of azimuthal harmonics is possible due to the decomposition of the total field **E**, **H** in terms of azimuthal harmonics (See SM). The scattering of the cylindrical harmonic with a certain azimuthal index $\sigma_m$ will contribute to the total SCS, which makes it possible to decompose the calculated spectra into azimuthal harmonics, as shown in Fig. 4 for the spectrum of a ring with aspect ratio $R_{out}/h=8.2$. In the calculations, we limited ourselves to the eighth-order harmonic, when a fairly good agreement was obtained between the spectra calculated by two different methods.

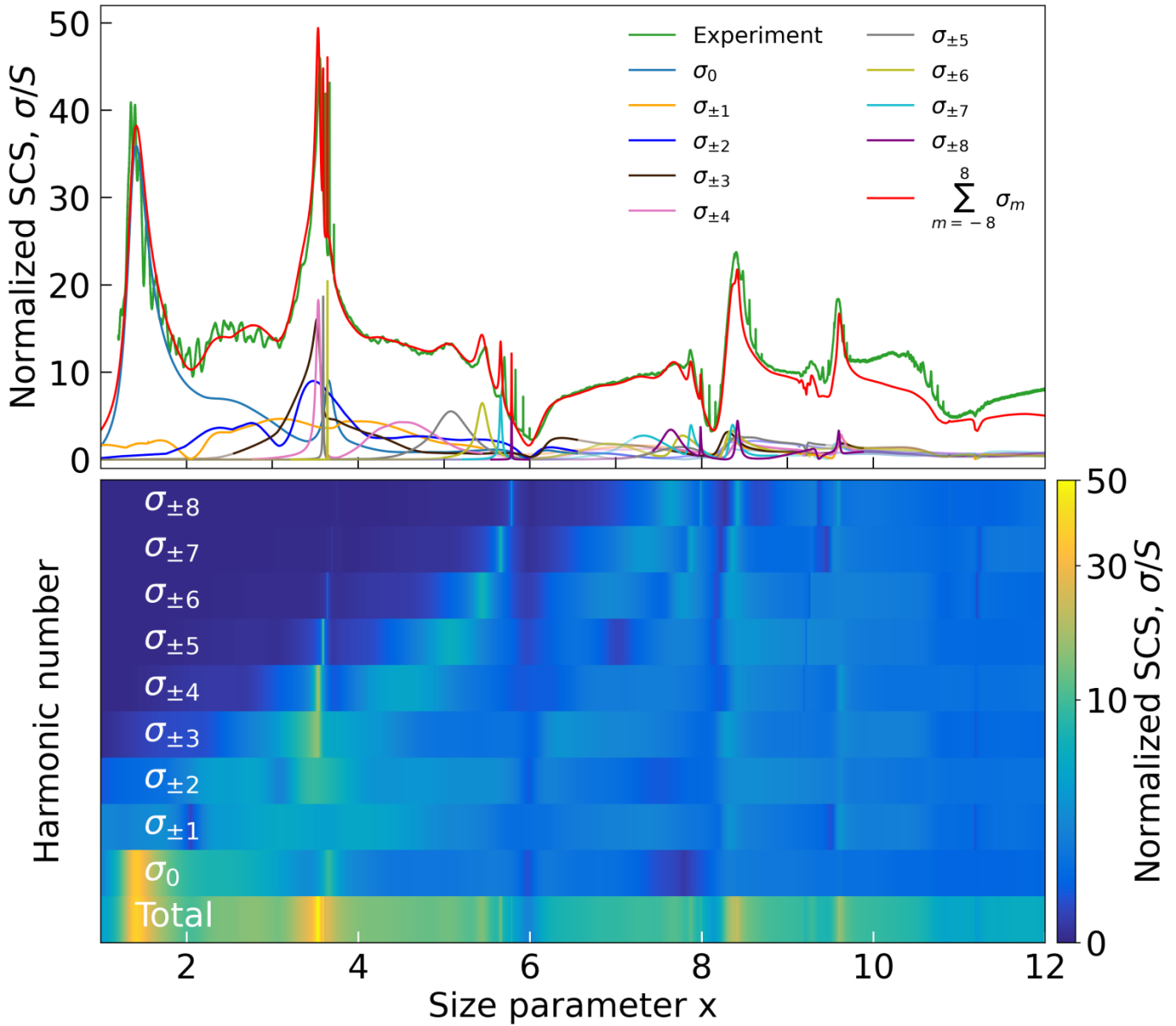

**Fig. 4.** Decomposition of the SCS spectrum in terms of azimuthal harmonics up to $m = 8$ inclusive. Upper panel: Experimentally measured SCS spectrum of RRs (green line) and spectra of partial harmonics up to the eighth inclusive, as well as their sum (red line). Bottom panel: The color map of partial harmonic intensities and their sums (Total). Aspect ratio $R_{out}/h = 8.2$, $R_{in}/R_{out}=0.81$, $\varepsilon=43$, $\varepsilon_{vac}=1$. $\sigma_{\pm m}= \sigma_{m}+ \sigma_{-m}$. $x = R_{out}\omega/c$.

The results obtained, presented in Fig. 4, make it possible to interpret a complex SCS spectrum. The first intense broad band at a frequency of $x \sim 1.4$ is determined by the intense dipole harmonic $m = 0$ with a weak addition of the harmonic $m = \pm 1$. The harmonic $m = \pm 2$ makes a significant contribution to the next broad and rather intense line at a frequency of $x \sim 2.55$. Further in the spectrum in the range of $x \in 3.3-3.8$, the first radial gallery is observed, which corresponds to fixed radial and axial indices $r = 1$, $z = 0$ and variable azimuthal indices $m$. The gallery begins with a first radial Fabry-Pérot-like resonance with a quasi-Lorentz contour, the width of which is determined by several harmonics at once, including $m = 0, \pm 2, \pm 3$. At the same time, on its high-frequency wing, there is a sequence of narrow equidistant lines from all harmonics $m = \pm 4, \pm 5, \ldots$ taken into account in our calculation. Next, the spectrum is determined by a second radial gallery with a radial index $r = 2$, an axial index $z = 0$, and a sequence of azimuthal equidistant harmonics $m$. As shown earlier [16], the substantially asymmetric shape of the broad Fabry-Perot-like resonance is determined by the Fano resonance [17, 18]. Due to the periodicity law of azimuthal Fano-harmonics $TE_m$ (See SM) their intensities turn to zero or close to zero at the same frequencies $\omega_{zero}$ ($x\sim6$ in Fig. 4).

**Experimental observation of radial and axial galleries.** An important part of this work was the experimental confirmation of the theoretical results, namely, the observation of individual radial and axial galleries and their shift in the frequency scale with a change in the geometric size of the RR. It is well known that there is no fundamental length scale in Maxwell's equations (SM). The use of centimeter-sized samples determined the microwave range of studies. Far-field SCS spectra of the RRs were measured at microwave frequencies in an anechoic chamber (SM). The technology for manufacturing RRs from $Ca_{0.67}La_{0.33})(Al_{0.33}Ti_{0.67})O_3$ ceramics is also described in the SM.

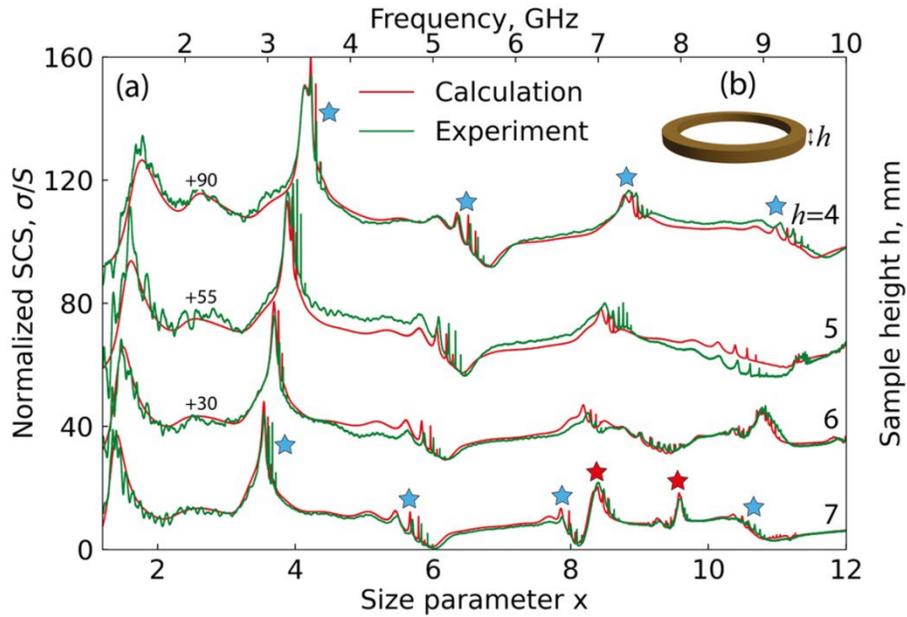

**Fig. 5.** (a) Experimentally measured (green lines) and calculated (red lines) SCS spectra of dielectric RRs as a function of the ring height $h$. Radial galleries are marked with blue stars and axial galleries with red stars. The length (along the middle line of the ring) and the width of all samples are constant: 326 mm and 11 mm, respectively. RRs with dielectric permittivity $\varepsilon=43$ are embedded in air, $\varepsilon_{air}=1$. TE-polarization. The spectra are shifted vertically by the values shown. $x = R_{out}\omega/c$. (b) Photograph of the examined sample.

Figure 5 shows experimentally measured and calculated SCS spectra of dielectric RRs with constant length and width, but variable heights of 4, 5, 6, and 7 mm. As discussed above, in the spectra of a 4 mm high sample at low frequencies, an intense dipole mode is observed ($m = 0$, spectral range 1.2-1.5 GHz), and another weaker peak ($m = \pm1$, ~2.2 GHz), after which galleries with two Lorentz ($r = 1, 3$) and two Fano ($r = 2, 4$) radial Fabry-Perot-like resonances alternate in the spectra. The Lorentz galleries are located in the spectral ranges 3-4 and 7-8 GHz, and the Fano galleries are located in the ranges 4.5-5.5 and 9-10 GHz. As the ring height increases, new axial galleries appear at high frequencies, which, as expected, demonstrate a strong frequency dependence on the sample height, repeating the previously observed dependence of the Fabry–Perot-like resonant frequencies in the spectra of cylindrical resonators of variable length [4].

**Conclusion.** The paper presents for the first time a complete picture of the photonic eigenmodes of a dielectric RR in the low-frequency region of the spectrum. The interference strongly influences the scattering pattern, making it dependent on the wavelength, showing both sharp narrow peaks and broad bands, including those with Fano contours. The use of ceramic samples with low material losses made it possible to experimentally observe narrow-band axial resonances following each broadband Fabry-Pérot-like resonance and thus confirm the structure of the spectrum consisting of individual galleries. An excellent agreement was obtained between numerical calculations and experimental data, including the dependence of the photonic resonant modes on all three indices - azimuthal ($m$), radial ($r$) and axial ($z$). By varying the width or height of the resonator, the radial (TE-like) or axial (TM-like) galleries can be independently shifted as desired, according to the requirements for specific applications.

**Funding.** Ministry of Science and Higher Education of the Russian Federation (Project 075-15-2021-589), Russian Science Foundation (Project No 23-12-00114).
**Acknowledgments.** AC, MS, NS (experiments and calculations) acknowledges the financing of the project No 23-12-00114, KS and ML (calculations and paper preparation) acknowledges the financing of the project 075-15-2021-589. The authors thank E. Nenasheva (Ceramics Co.Ltd., St. Petersburg) for providing samples for measurements.
**Disclosures.** The authors declare no conflicts of interest.
**Data availability.** Data underlying the results presented in this paper are not publicly available at this time but may be obtained from the authors upon reasonable request.
**Supplemental document.** See Supplement 1 for supporting content.

# Supplementary materials

## Optical galleries of dielectric rings: antipode to whispering gallery modes

A. P. Chetverikova, M. S. Sidorenko, K. B. Samusev, M. F. Limonov, N. S. Solodovchenko

### A. Scattering spectrum decomposition in terms of azimuthal harmonics

Determination of azimuthal harmonics is possible due to the decomposition of the total field E, H in terms of azimuthal harmonics:

$$\mathbf{E} = \sum_{m=-\infty}^{\infty} = \mathbf{E}_m exp(-im\phi), \quad \mathbf{H} = \sum_{m=-\infty}^{\infty} = \mathbf{H}_m exp(-im\phi) \tag{A.1}$$

The total electric field is represented as the sum of the incident $E_{back}$ and scattered $E_{sc}$ fields

$$E = E_{back} + E_{sc}. \tag{A.2}$$

The incident field in this problem is a plane wave, the electric vector of which is perpendicular to the RR axis (z axis), and the wave vector is directed along the x axis (TE polarization):

$$\mathbf{E} = E_0 e^{-i\mathbf{kr}} \cdot \mathbf{i}_y. \tag{A.3}$$

A plane wave can be decomposed in terms of cylindrical Bessel functions

$$E_\rho = \frac{E_0}{2} \sum_{m=-\infty}^{\infty} (-i)^{m+1} e^{-im\phi} \left[ J_{m+1}(k_x\rho) - J_{m-1}(k_x\rho) \right],$$

$$E_\phi = -\frac{E_0}{2} \sum_{m=-\infty}^{\infty} (-i)^m e^{-im\phi} \left[ J_{m+1}(k_x\rho) + J_{m-1}(k_x\rho) \right],$$

$$E_z = 0. \tag{A.4}$$

The scattering cross section is defined as

$$\sigma = \frac{c}{8\pi} \int Re\left(\mathbf{E} \times \mathbf{H}^*\right) d\mathbf{S} \Big/ I_0, \tag{A.5}$$

and substitution of decompositions (A.1) into expression (A.5) gives

$$\sigma = \frac{c}{8\pi} \sum_{m=-\infty}^{\infty} \int Re\left(\mathbf{E}_m \times \mathbf{H}_m^*\right) dS \Bigg/ I_0 = \sum_{m=-\infty}^{\infty} \sigma_m.$$
(A.6)

Integration over the angle $\varphi$ gives the delta function $\delta_{mn}$, therefore, the scattering cross section $\sigma$ is the sum of the partial cross sections $\sigma_m$ with different azimuthal indices $m$. Thus, the scattering of the cylindrical harmonic (1.6-1.8) with a certain azimuthal index $m$ will contribute to the total scattering cross section. Thus, the scattering of the cylindrical harmonic (1.6-1.8) with a certain azimuthal index $m$ will contribute to the total scattering cross section, which makes it possible to decompose the calculated spectra presented in Fig. 2 (A) into azimuthal harmonics, as shown in Fig. 4 for the spectrum of a ring 7 mm thick. In the calculations, we limited ourselves to the eights order, when a good agreement was obtained between the spectra calculated by two different methods. In our situation, the azimuth harmonic of the eighth order is sufficient.

To calculate partial scattering cross section in Comsol, we used "2D axisymmetric space", in which a geometric section of a ring surrounded by a PML layer is projected (Figure S1). By choosing a certain value of the azumital harmonic m (for example, m=0) in formula (A.4), one can obtain an explicit form of the incident field (E$\rho$,E$\varphi$) and solve the scattering problem. To calculate the scattering cross section, the energy of the scattered field is collected at the inner boundary of the PML (A. 6). To calculate the eigenvalues of the RR, we used a geometry similar to the scattering problem, but in this case there was no incident field.

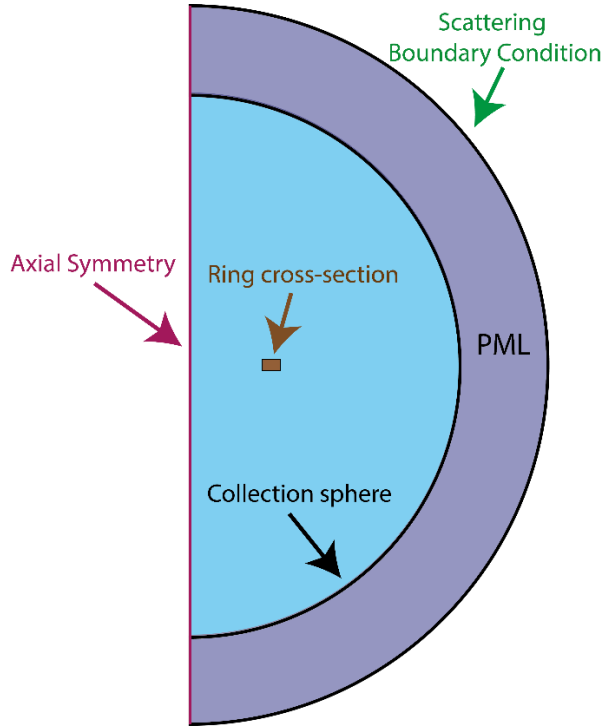

**Fig. S1**: Geometry of the problem for calculating partial scattering cross section.

**B. Dependence of the SCS spectra in the wavelength scale.**

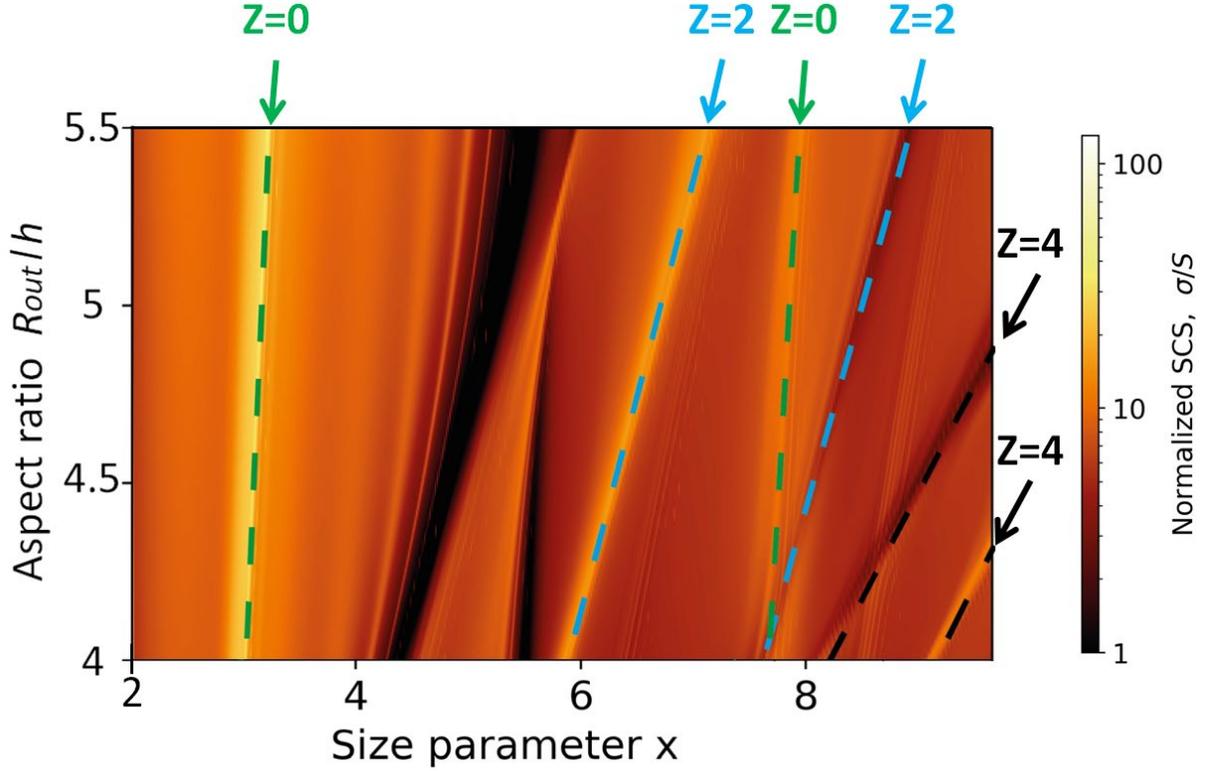

**Fig. S2**. Dependence of the normalized SCS spectra of dielectric RRs on ring height $h$ in the wavelength scale. The spectra were calculated with a step of $\Delta(R_{out}/h) = 0.01$ along the ring height. For understanding, it should be noted that the linear dependences in Fig. S1 are observed in the axes ($R_{out}/h$, $R_{out}\omega/c$), which is equivalent to the observation of linear dependences in the inverted axes ($h$, $\lambda$).
Arrows and dashed lines marked with different colors show linear dependences of the wavelength on the ring height for Fabry- Pérot-like resonances with axial indices z=0, 2, 4.
TE polarization, $\varepsilon = 43$, $\varepsilon_{vac} = 1$. The size parameter $x = R_{out}\,\omega/c$.

## C. Determination of the Fano parameter $q$.

To find the Fano parameter, we will use the solution of the Mie problem for a two-layer medium. A similar problem was solved for the cylinder [1]

$$\left(\frac{d_{m2}}{\sqrt{\varepsilon}^2}\frac{\frac{\partial}{\partial x_2}J_m(\sqrt{\varepsilon}x_2)}{\frac{\partial}{\partial x_2}H_m^1(x_2)} + \frac{a_{m2}}{\sqrt{\varepsilon}^2}\frac{\frac{\partial}{\partial x_2}H_m^1(\sqrt{\varepsilon}x_2)}{\frac{\partial}{\partial x_2}H_m^1(x_2)}\right) + \left(-\frac{\frac{\partial}{\partial x_2}J_m(x_2)}{\frac{\partial}{\partial x_2}H_m^1(x_2)}\right) = a_{m3} \tag{C.1}$$

Here $J_m(x)$ is a Bessel function, $H_m(x)$ a Hankel function of the first kind, and $x_1 = kR_{in}$, $x_2 = kR_{out}$. We assume that perfect electric conductor (PEC) ring is equivalent to PEC cylinder with same outer radius. Applying the PEC limit, it could be discovered that scattering trends to:

$$a_{m3} \to a_m^{(PEC)} = \frac{\frac{\partial}{\partial x_2}J_m(x_2)}{\frac{\partial}{\partial x_2}H_m^1(x_2)} = \frac{J_m'(x_2)}{J_m'(x_2) + iY_m'(x_2)} \tag{C.2}$$

With this result, we can describe a phase relation for dielectric ring as for dielectric cylinder, in next way:

$$a_m = a_m^{(PEC)} - \frac{1}{\sqrt{\varepsilon}}\frac{J_m'(\sqrt{\varepsilon}x_2)}{H_m'(x_2)}d_m \tag{C.3}$$

From the other hand, the $a_n$ coefficient could be described as:

$$a_m = \frac{1}{1 + i\cot(\Delta_m^a)} \tag{C.4}$$

Now, we can analytically describe the phase relation for ring in this way:

$$\Delta_m^a = \Delta_m^{a,PEC} + \Delta_m^{a,res} \tag{C.5}$$

Taking a cot for this equation we obtain:

$$\cot(\Delta_m^a) = \cot(\Delta_m^{a,PEC} + \Delta_m^{a,res}) = \frac{\cot(\Delta_m^{a,PEC})\cot(\Delta_m^{a,res})}{\cot(\Delta_m^{a,PEC}) + \cot(\Delta_m^{a,res})} \tag{C.6}$$

We already know relation for $\Delta^{a,PEC}$ from equation (C.2), which now it is convenient to write in next view:

$$\cot(\Delta_m^{a,PEC}) = -\frac{Y_m'(x_2)}{J_m'(x_2)} \tag{C.7}$$

With implementation of new notations:

$$q = \cot(\Delta_m^{a,PEC}), \xi = \cot(\Delta_m^{a,res}) \tag{C.8}$$

and substitution them into equation (C.7) and then into (C.4), we finally obtain scattering intensity:

$$|a_m|^2 = \frac{(\xi + q)^2}{(\xi^2 + 1)(q^2 + 1)} \tag{C.9}$$

In the fig.2(a), we show Fano parameter q, according to (C.8), and fig.2(b) describes background field, which could be found with using:

$$BG = \left| -\frac{\frac{\partial}{\partial x_2} J_m(x_2)}{\frac{\partial}{\partial x_2} H_m^1(x_2)} \right|^2 \qquad \text{(C.10)}$$

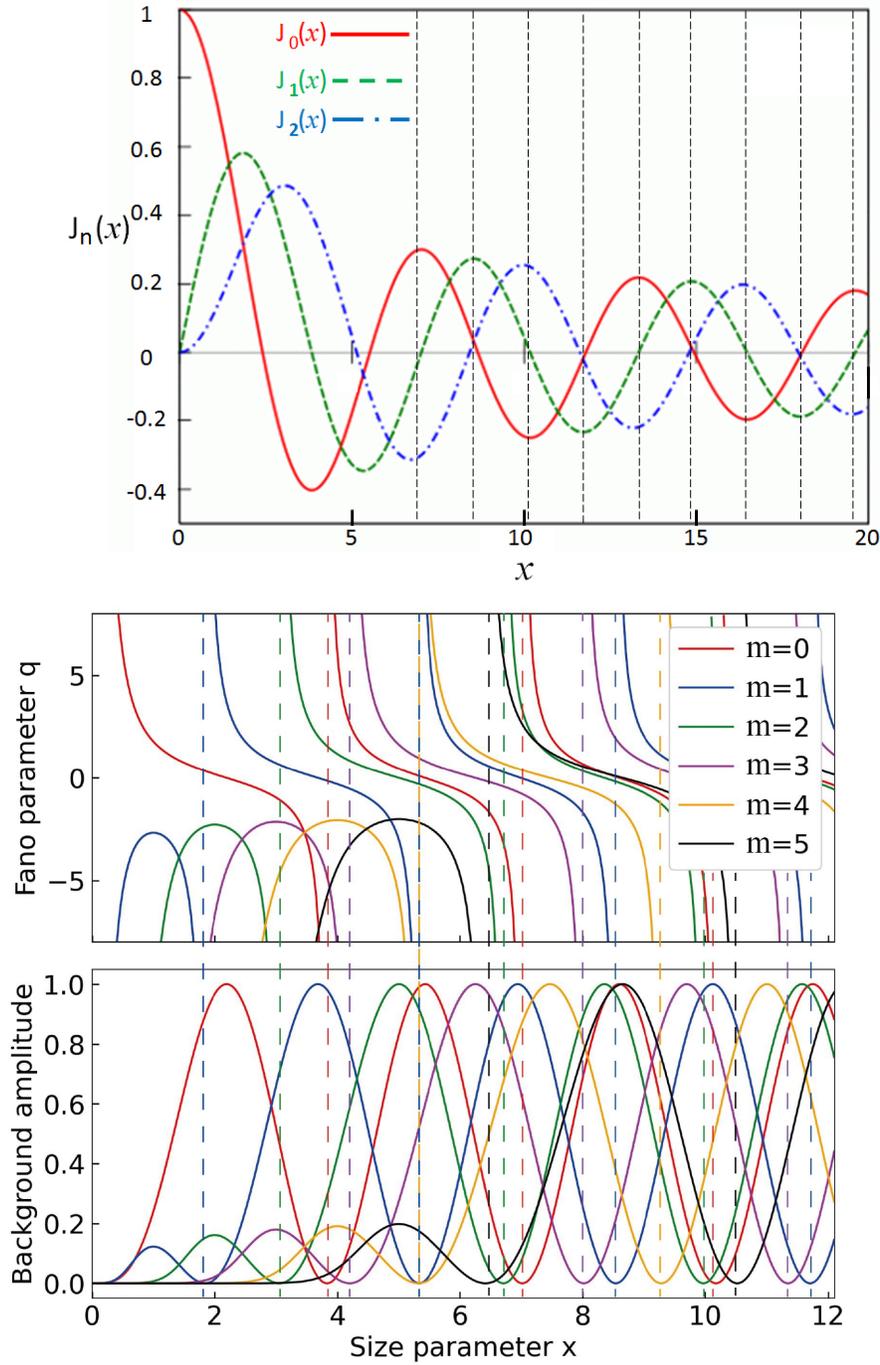

**Fig. S4.**
a) Bessel functions of the first kind, Jn(x), for integer orders n = 0, 1, 2. The vertical dashed lines show that extremes (maxima, minima, zeros) of Bessel functions with different n are observed for the same values of the argument, periodically repeating after the first cycle, x > 5.
b) Fano parameter q depending on size parameter x obtained by the formula (C.8)

c) Background amplitude constructed by the formula (C.10).

### D. Scaling properties of maxwell's equations.

The wave equation with an explicit dependence of the parameters on the position has the form:

$$\nabla \times \frac{1}{\mu(\vec{r})} \nabla \times \vec{E}(\vec{r}) = \omega^2 \mu_0 \varepsilon_0 \varepsilon(\vec{r}) \vec{E}(\vec{r}) \tag{D.1}$$

Next, we scale the dimensions by the $R_{out}$ ring parameter:

$$(R_{out}\nabla) \times \frac{1}{\mu(\vec{r}/R_{out})} (R_{out}\nabla) \times \vec{E}(\vec{r}/R_{out}) = \omega^2 \mu_0 \varepsilon_0 \varepsilon(\vec{r}/R_{out}) \vec{E}(\vec{r}/R_{out}) \tag{D.2}$$

Simplifying equation (5), we transfer the scale factor $R_{out}$ into one bracket with the frequency $\omega$ and introduce the parameter $\vec{l} = \vec{r}/R_{out}$:

$$\nabla \times \frac{1}{\mu(\vec{l})} \nabla \times \vec{E}(\vec{l}) = (\frac{\omega}{R_{out}})^2 \mu_0 \varepsilon_0 \varepsilon(\vec{l}) \vec{E}(\vec{l}). \tag{D.3}$$

The frequency scaling effect is just a size scaling, and by choosing a spectral interval convenient for the experiment, we only need to choose the correct sample size and, conversely, for a sample of a certain size, the corresponding spectral interval is selected.

E. **Experimrntal setup.**

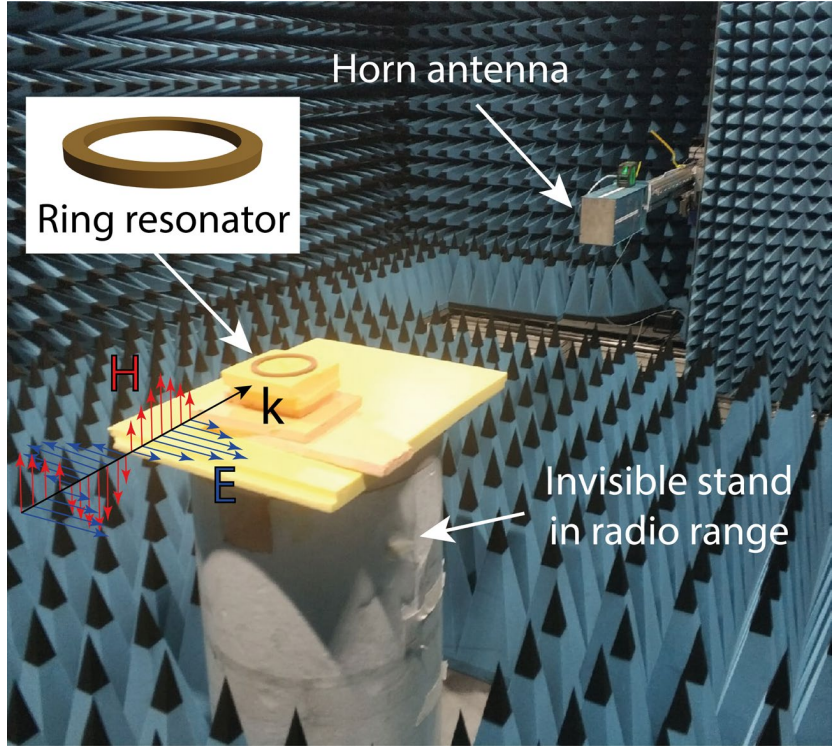

**Fig. S5**. Experimental setup

Experimental studies were carried out in an anechoic chamber. A pair of wideband horn antennas (TRIM 0.75 to 18 GHz; DR) were positioned facing each other at a distance of 4 m with the sample placed at the midpoint. The measurement used a two ports vector network analyzer (VNA; Agilent E8362C) transmitting a continuous wave. The first antenna was connected to the first port of the VNA and provided a near plane-wave excitation in the frequency range of 0.8 to 5 GHz. The second horn antenna connected to the second port of the VNA was employed as a receiver. The frequency range of 0.8 to 5 GHz was swept using 60001 frequency points. Eight such sweeps were averaged for each of the sample measurement, background measurement, and calibration measurement. Calibration measurements were performed by our colleagues using a metal sphere with the radius of 7.5 mm [2].

The forward scattering amplitude is obtained from the measured transmission coefficient between two antennas as $S_{21}/S_{21,0}$, where $S_{21,0}$ is the free space transmission coefficient, and $S_{21}$ is measured in the presence of the ring. According to the optical theorem, the imaginary part of the forward scattering amplitude is proportional to the scattering cross-section.

$$\sigma \sim \mathrm{Im}\left(\frac{S_{21}}{S_{21,0}} - 1\right)/k \qquad (\text{E.1})$$

F. **Sample fabrication technology.**

Sample fabrication technology is described in detail in our previous article [N.Solodovechenko, M. Sidorenko, T. Seidov, I. Popov, E. Nenasheva, K. Samusev, and M. Limonov, "Cascades of Fano resonances in light scattering by dielectric particles", Materials Today, 60, 69 (2022)]. Our studies were carried out on ceramic samples $(Ca_{0,67}La_{0,33})(Al_{0,33}Ti_{0,67})O_3$. Ceramic samples were prepared using solid state ceramic route. As the starting reagents served $CaCO_3$ (99,5wt%, average grain size D=2µm), $La_2O_3$(>99,85wt%, D=1µm), $Al_2O_3$(>99,5wt%, D=1µm) and $TiO_2$(>99,3wt%, D=0,5µm). After milling in a vibration mill for three hours the mixture was calcined in air at $1400^0$C for four hours, then the calcined powder was re-milled by ball milling for four hours to a grain size (0,5-1)µm. Samples of rings of geometric sizes required for experiments were obtained by hydraulic pressing. The prepared samples were sintered in air within the temperature range of 1540°C (3h) in a chamber electric furnace until zero water absorbance and porosity less than 4% was obtained.